\documentclass{aa}  

\usepackage{graphicx}
\usepackage{txfonts}

\usepackage{color,soul}
\usepackage{multirow}

\newcommand{\sneia}{{SNe~Ia}\xspace}
\newcommand{\snia}{{SN~Ia}\xspace}
\newcommand{\abc}{{iPTF16abc}\xspace}
\newcommand{\host}{{NGC~5221}\xspace}
\newcommand{\los}{{line-of-sight}\xspace}
\newcommand{\loss}{{lines-of-sight}\xspace}
\newcommand{\naid}{{Na~I~D}\xspace}  
\newcommand{\cahk}{{Ca~II~H\&K}\xspace}
\newcommand{\ki}{{K~I}\xspace}  
\newcommand{\nai}{{Na~I}\xspace}  
\newcommand{\caii}{{Ca~II}\xspace}
\newcommand{\hi}{{H~I}\xspace}

\newcommand{\swift}{{\em Swift}\xspace}
 
\begin{document} 

   \title{
   Probing gas and dust in the tidal tail of \host with the type Ia supernova \abc
  \thanks{Based on observations collected at the European Organisation for Astronomical Research in the Southern Hemisphere under ESO DDT programme 297.D-5005(A), P.I. Ferretti.}}

   \author{R.~Ferretti  \inst{1}
          \and
          R.~Amanullah  \inst{1}
          \and
          A.~Goobar   \inst{1}
          \and          
          T.~Petrushevska \inst{1}
          \and
          S.~Borthakur \inst{2}
          \and
          M.~Bulla \inst{1}
          \and
          O.~Fox \inst{3}
          \and     
          E.~Freeland \inst{4}
          \and
          C.~Fremling \inst{4}
          \and 
          L.~Hangard \inst{1}
          \and
          M.~Hayes \inst{4}
          }

   \institute{
   	Department of Physics, The Oskar Klein Centre, Stockholm University, Albanova, SE 106 92 Stockholm, Sweden\\
	 \email{raphael.ferretti@fysik.su.se}
        \and
        Department of Physics \& Astronomy, john Hopkins University, 3701 San Martin Drive Baltimore, MD 21218\
        \and
	Space Telescope Science Institute, 3700 San Martin Drive, Baltimore, MD 21218, USA\
	\and
	Department of Astronomy, The Oskar Klein Center, Stockholm University, Albanova, SE 10691 Stockholm, Sweden\
             }

   \date{Received ... ; accepted ... } 

 
  \abstract
   {Type Ia supernovae (\sneia) can be used to address numerous questions in astrophysics and cosmology.
   Due to their well known spectral and photometric properties, \sneia are well suited to study gas and dust along 
   the \loss to the explosions. For example, narrow \naid and \cahk absorption lines can be studied easily, 
   because of the well-defined spectral continuum of \sneia around these features.
   }
   {
   We study the gas and dust along the \los to \abc, which occurred in an unusual location,
   in a tidal arm, $80$~kpc from centre of the galaxy \host. 
   }
   {Using a time-series of high-resolution spectra, we examine narrow \naid and \cahk absorption features 
   for variations in time, which would be indicative for circumstellar (CS) matter.
   Furthermore, we take advantage of the well known photometric properties of \sneia to determine reddening due to 
   dust along the \los.
   }
   {From the lack of variations in \naid and \cahk, we determine that none of the detected absorption features originate from the 
   CS medium of \abc.
   While the \naid and \cahk absorption is found to be optically thick, a negligible amount of reddening points to 
   a small column of interstellar dust.
   }
   {We find that the gas along the \los to \abc is typical of what might be found in the interstellar medium (ISM)
   within a galaxy.
   It suggests that we are observing gas that has been tidally stripped during an interaction of \host with one of its 
   neighbouring galaxies in the past $\sim10^9$ years.
   In the future, the gas clouds could become the locations of star formation. 
   On a longer time scale, the clouds might diffuse, enriching the  
   circum-galactic medium (CGM) with metals.
   The gas profile along the \los should be useful for future studies of the dynamics of the galaxy group containing \host. 
   }

   \keywords{
     supernovae: individual: \abc --        
     Galaxies: individual: \host --
     Galaxies: ISM, interactions
               }

   \maketitle
%

\section{Introduction}

Type Ia supernovae (\sneia) have such uniform properties that they are frequently described 
as "standard candles", which, among other applications, enables detailed measurements
in cosmology \citep[see for instance][]{2011ARNPS..61..251G}.
The well-known brightness and spectral energy distribution (SED) of \sneia
can also be used to study gas and dust along the \loss.

Due to the well defined continuum of \sneia, typical absorption features of the interstellar medium (ISM)
of their host galaxies, such as \naid and \cahk, can frequently be identified.
Furthermore, diffuse interstellar bands (DIBs), which are largely 
of unknown chemical origin \citep{2001AcSpe..57..615S,2006ARA&amp;A..44..367S},
are sometimes detected in the spectra \citep[e.g.][]{2005A&amp;A...429..559S}.
One can also take advantage of the standard candle properties of \sneia to study dust in the ISM.
Comparing the lightcurves of a given \snia with an SED template, 
extinction and reddening can be determined 
to study the dust properties along the \los.

For example, the nearest \snia in modern times, SN~2014J, \citep{2014ApJ...784L..12G},
has been used to study details of the ISM composition and structure 
of M82 \citep{2014ApJ...792..106W,2015ApJ...799..197R,2017ApJ...834...60Y}.
Interestingly, an unusual extinction curve was identified which possibly points 
to very small dust grains in the ISM or dust in the circumstellar (CS) medium of the 
supernova \citep{2014ApJ...788L..21A, 2014MNRAS.443.2887F}. 

\sneia can be used to study a large variety of \loss. 
Due to the large delay times of $\approx10^{8}$--$10^9$ years between 
the formation of a \snia progenitor system 
and the explosion \citep{2012MNRAS.426.3282M},
they are known to occur almost anywhere in their host galaxies.
An extreme example is PTF10ops, which was located at a projected distance of $148$~kpc 
from the nearest possible host galaxy \citep{2011MNRAS.418..747M}.

It is thus possible for \sneia to occur in regions with low stellar density, such as 
tidally stripped spiral arms of galaxies.
Tidal arms are thought to be the result of close interactions of
neighbouring galaxies \citep[as shown in the simulations by][]{1972ApJ...178..623T}.
\sneia occurring in such locations can be used to study the interstellar gas and dust properties of regions 
that have been affected by gravitational interactions.
The ISM in tidal arms can be the location of future star formation \citep{1990AJ.....99..497S} 
and enrich the circumgalactic medium (CGM) with metals.

To date no \snia located in a tidal arm has been studied in detail,
although at least one case is known \citep[SN~2013de,][]{2013CBET.3555....1D,2013ATel.5097....1D}.
Here we investigate \abc, a \snia which occurred in the tidal arm of
its host galaxy, \host.
Due to the large displacement of \abc from its host, the supernova was a 
good candidate to search for the presence of CS gas via photoionisation
using the method described in \citet{2016A&amp;A...592A..40F}.
Furthermore, the obtained high-resolution spectra and photometry could be used to 
study the spiral arm where the supernova is located.

In the following, we present the discovery and observations of \abc (Section~\ref{sec:disc}).
We then describe the narrow absorption features detected in high-resolution spectra, as well as the 
photometric reddening of \abc (Section~\ref{sec:hires}).
We further show that the deep absorption features cannot be due to CS gas surrounding the supernova, 
but are part of the tidally stripped ISM of \host (Section~\ref{sec:cs}).
Finally, we compare the \los with \host and discuss the features in the context 
of the galaxy group in which \host is located (Sections~\ref{sec:host} and~\ref{sec:con}).


\section{Discovery and Observations of \abc}
\label{sec:disc}

\begin{figure*}
 \centering
 \resizebox{\hsize}{!}{\includegraphics{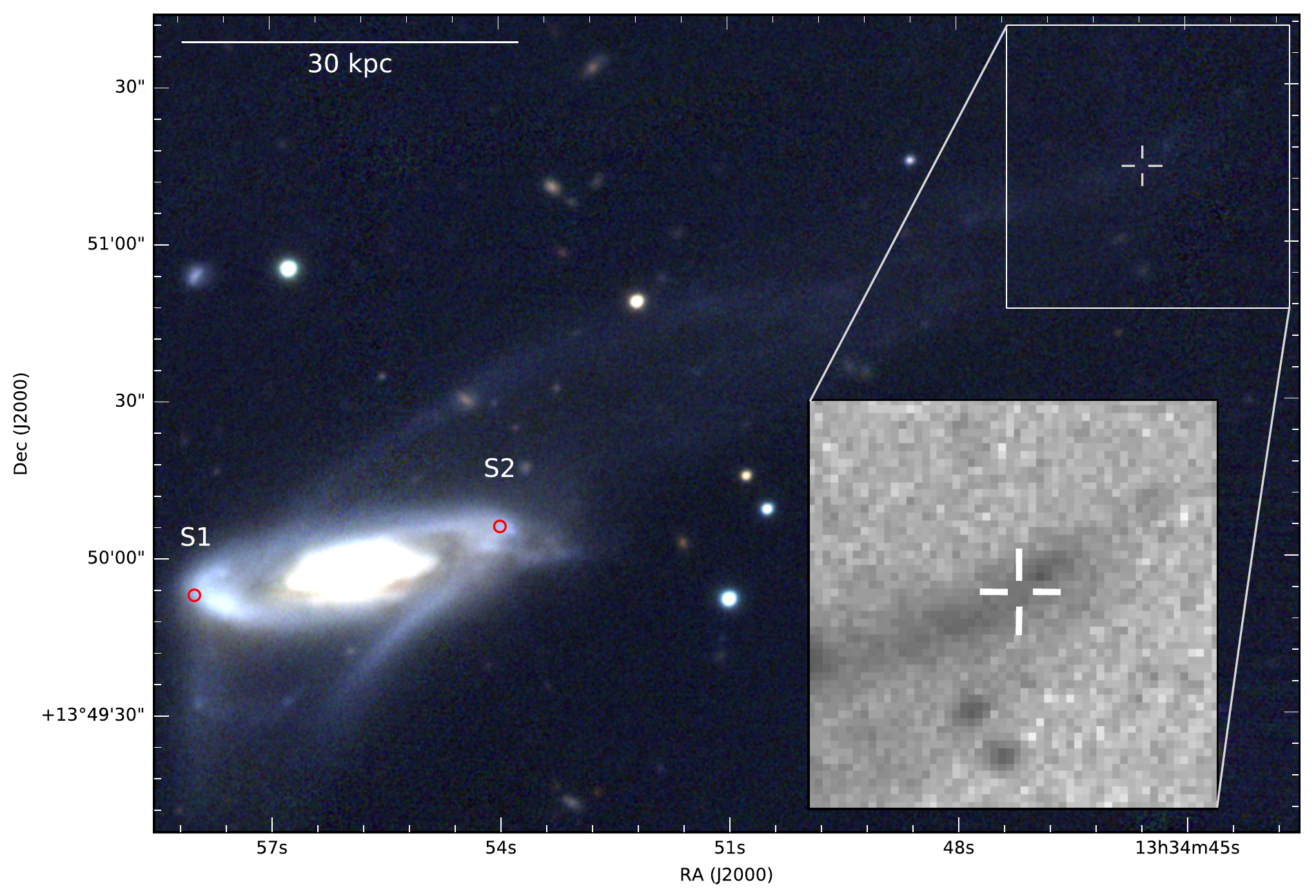}}
     \caption{The location of \abc, marked in the inset, with respect to its host galaxy \host. 
     The tidal tail is visible spanning the space between the galaxy and the supernova location.
     Coordinates at which SDSS spectra of \host are available are marked and labeled S1 and S2.
     The main image is a colour composite of the publicly available PanSTARRS data \citep{2016arXiv161205243F}.
     The inset is a deep stack of pre-explosion iPTF P48 R-band images. 
       \label{fig:field}}
 \end{figure*}

The intermediate Palomar Transient Factory (iPTF) first reported the discovery of \abc \citep{2016ATel.8907....1M},
located $170"$ from the galaxy \host at R.A., Dec = 13:34:45.492 +13:51:14.30 (J2000),
a \los that has low Milky Way reddening with $E(B-V)_{MW}=0.028$~mag \citep{2011ApJ...737..103S}.
A colour composite image of \host and the location of \abc is shown in Figure~\ref{fig:field}.
The supernova was discovered on 4.4~April~2016, 
while the first detection was on 3.4~April 
and the last non-detection occurred on 2.4~April 
(UT dates are used throughout the paper).

A classification spectrum of \abc was obtained with the 
DeVeny spectrograph on the Discovery Channel Telescope (DCT) \citep{2016ATel.8909....1C} 
on 5.3~April, 
which determined it to be a \snia, 
possibly belonging to the subclasses resembling SNe 1991T or 1999aa.
A spectrum obtained with Gemini N+GMOS on 5.4~April 
revealed the presence of deep \naid and \cahk absorption features. 
In response, extensive photometric and spectroscopic follow-up were triggered 
(Miller et al, 2017; submitted).

As part of the follow-up of \abc, we received Director's Discretionary Time 
at the European Organisation for Astronomical Research in the 
Southern Hemisphere (ESO DDT programme 297.D-5005(A), P.I. Ferretti).
The spectra taken with
Xshooter \citep[$R\simeq7500$,][]{2011A&amp;A...536A.105V} and the
Ultraviolet and Visual Echelle Spectrograph  \citep[UVES; $R\simeq50000$][]{2000SPIE.4008..534D}  
both on Kueyen (UT2) $8.2$~m unit at the Very Large Telescope (VLT) 
are summarised in Table~\ref{tab:spec}.
The Xshooter  and UVES spectra were reduced using the REFLEX (ESOREX) reduction pipeline
provided by ESO \citep{2010SPIE.7737E..28M}. 
Where necessary, telluric features were corrected for using Molecfit \citep{2015A&amp;A...576A..77S,2015A&amp;A...576A..78K}.

We analyse the photometry of \abc presented in Miller et al. (2017; submitted).
From the Palomar Observatory 48-inch telescope (P48) we use $g$- and $R$-band photometry 
which was obtained as part of the regular iPTF survey.
Further we use $gr$- and $i$-band photometry of the Palomar Observatory 60-inch telescope (P60) and
$BVgr$- and $i$-band from the Las Cumbres Observatory (LCO).
Ultraviolet (UV) photometry from the Ultraviolet/Optical Telescope \citep[UVOT;][]{2005SSRv..120...95R} 
on the \swift spacecraft \citep{2004ApJ...611.1005G} {\it uvm2}-filter 
and $riZYJ$- and $H$-band with the Reionisation and Transient Infrared/Optical Project \citep[RATIR;][]{2012SPIE.8453E..1OF}
was obtained and used for this analysis.

    \begin{table*}
       	\centering
  	\begin{tabular}{l c l@{\,}r c c c}
    	\hline\hline
    	{\small Instrument} & {\small MJD} & \multicolumn{2}{c}{\small UT Date}  &
	{\small Exp. time} &   {\small Phase} 
	& {\small Set-up}\\
	& & \multicolumn{2}{c}{\small } & {\small (s)} & {\small (days)}\\
    	\hline
	{\small Xshooter} & {\small 57,492.2} & {\small Apr.}&{\small 14.2} & {\small $1755$} & {\small -6.6} 
	& {\small UVB 1.0"/VIS 0.9"/NIR 0.9"}\\
	{\small UVES} & {\small 57,495.2} & {\small Apr.}&{\small 17.2} & {\small $2 \times 1700$} & {\small -3.6} 
	& {\small 0.8" DIC1 390+580}\\
	{\small UVES} & {\small 57,503.1} & {\small Apr.}&{\small 25.1} & {\small $3450$} & {\small 4.3} 
	& {\small 0.8" DIC1 390+580}\\
	{\small Xshooter} & {\small 57,520.0} & {\small May}&{\small 12.0} & {\small $1829$} & {\small 21.2} 
	& {\small UVB 1.0"/VIS 0.9"/NIR 0.9"}\\
	\hline\hline
	\multicolumn{5}{l}{\small {}}\\
  	\end{tabular}
  	\caption{Mid- and high-resolution spectra obtained
	with stretch-corrected phases in the rest frame with respect to B-band maximum.
	\label{tab:spec}}
\end{table*}

\section{Absorption features and reddening}
\label{sec:hires}

    \begin{table}
       	\centering
  	\begin{tabular}{c l@{\,}r l@{\,}r }
    	\hline\hline
    	{\small $v$} & \multicolumn{2}{c}{\small $b$}  & \multicolumn{2}{c}{\small ${\rm log}_{10}\{N_{\rm \caii}\}$} \\
	{\small (km s$^{-1}$)} & \multicolumn{2}{c}{\small (km s$^{-1}$)}  & \multicolumn{2}{c}{\small (cm$^{-2}$)} \\
    	\hline
	{\small -88.} & {\small 3.1} & {\small $\pm$ 1.3} & {\small 11.05} & {\small $\pm$ 0.09}\\
	{\small -80.} & {\small 5.0} & {\small $\pm$ 0.4} & {\small 11.93} & {\small $\pm$ 0.02}\\
	{\small -58.} & {\small 5.0} & {\small $\pm$ 0.3} & {\small 12.06} & {\small $\pm$ 0.02}\\
	{\small -52.} & {\small 3.6} & {\small $\pm$ 0.3} & {\small 12.24} & {\small $\pm$ 0.03}\\
	{\small -43.} & {\small 3.1} & {\small $\pm$ 1.1} & {\small 11.20} & {\small $\pm$ 0.08}\\	
	\hline\hline
	\multicolumn{5}{l}{\small {}}\\
  	\end{tabular}
  	\caption{Average Voigt profile parameters of \cahk in the two UVES spectra.
	\label{tab:vp}}
\end{table}

\begin{figure*}
   \centering
   	\resizebox{\hsize}{!}{\includegraphics{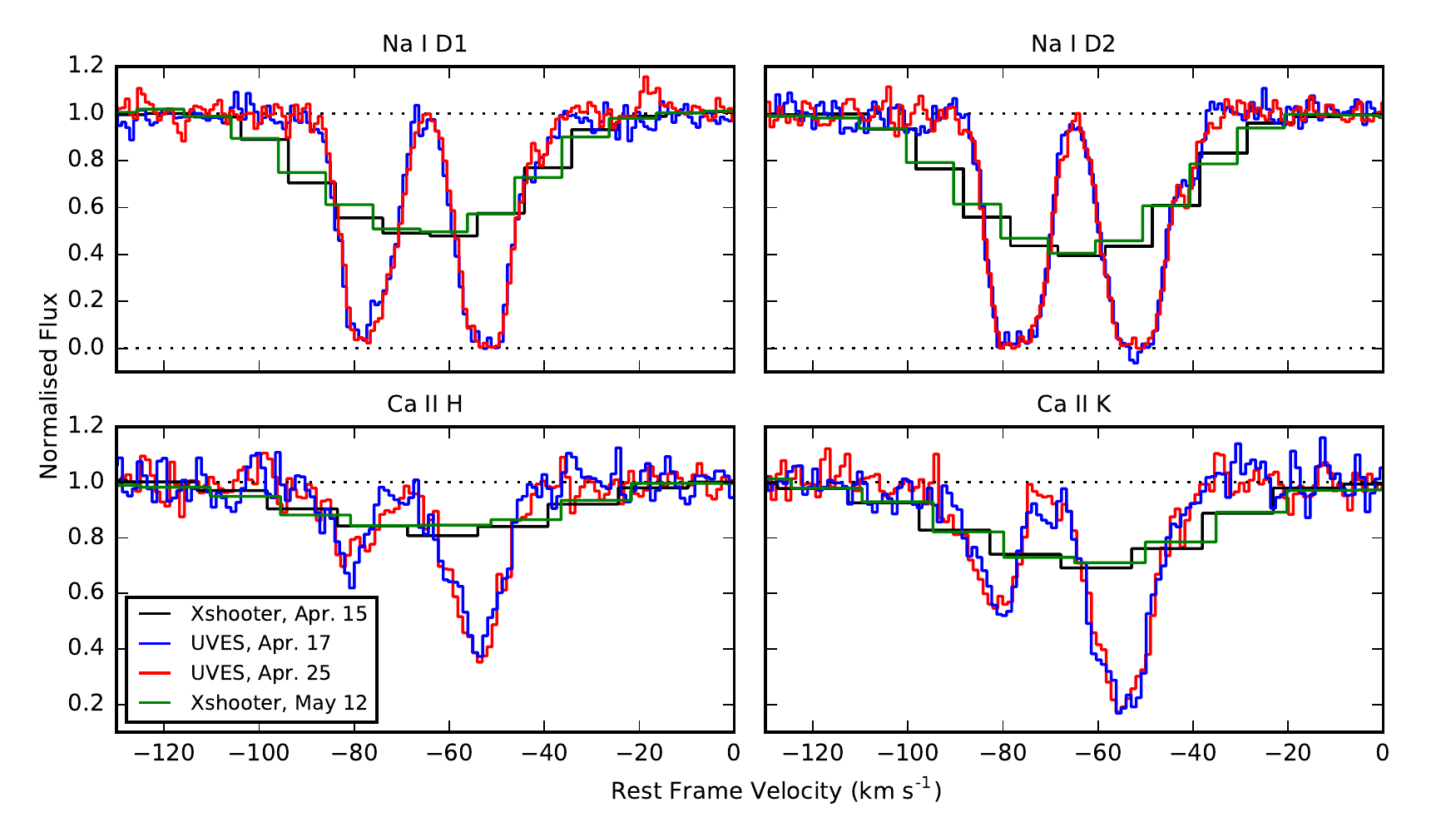}}
   \caption{The Na~I~D doublet and Ca~II~H\&K of iPTF16abc by UVES and Xshooter. 
   In the UVES spectra, two distinct groups of absorption features with similar velocities are visible, 
   while they are unresolved with Xshooter. 
   The substructure of these groups is difficult to further discern due to the optical thickness of the absorbers.
 }
         \label{fig:naid}
    \end{figure*}

Deep well-defined narrow absorption lines corresponding to \naid and \cahk at the rest frame 
of \host \citep[$z=0.0234$,][]{2015MNRAS.447.1531C}
are evident in all Xshooter and UVES spectra obtained.
We detect two distinct groups of absorbers at different radial velocities in the UVES spectra.
The absorption line profiles can be seen in Figure~\ref{fig:naid}, 
where the two deep feature groups are centred at a radial velocity $v_r\approx-77$ and $-51$~km~s$^{-1}$,
with respect to the host galaxy rest frame.

Both groups of absorption line features are visibly asymmetric, 
indicating that the absorption features are composed of several blended components.
The \naid lines are too saturated and blended to determine any further substructure with confidence.  
Therefore, only \nai column density upper limits can be determined.
Using VPFIT\footnote{\url{http://www.ast.cam.ac.uk/~rfc/vpfit.html}},
we fit Voigt profiles to \cahk. In Table~\ref{tab:vp}, we present the average Doppler width ($b$) and 
column densities ($N$) of a 5-component fit.
Since all the features are strongly blended, it cannot be excluded that there are 
more unresolved components contributing to the profile. 

\begin{figure}
   \centering
   	\resizebox{\hsize}{!}{\includegraphics{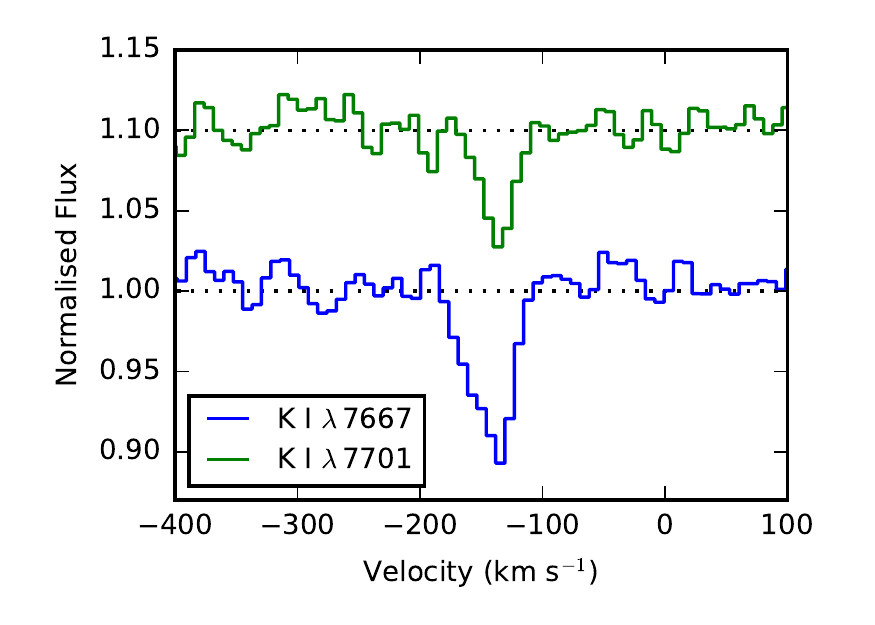}}
   \caption{\ki at $\lambda\lambda 7667$ and $7701$ from the 4~April Xshooter spectrum. 
   The spectra are normalised and corrected for telluric features.
 }
         \label{fig:ki}
    \end{figure}

We further detect \ki at $\lambda\lambda 7667$ and $7701$ in the Xshooter spectra.
Unfortunately, \ki is not covered by UVES with the chosen set-up.
In Figure~\ref{fig:ki}, the \ki lines from the spectrum of 4~April is shown, from which
we measure EWs~$=112\pm9$ and $71\pm6$~m\AA, respectively.
While the \ki is also visible in the spectrum from 12~May, several noisy features that do not correspond to telluric lines 
make precise EW measurements difficult.

We searched all spectra for traces of DIBs as well as CH and CH$+$.
We found a feature in both UVES and Xshooter 
spectra which seemingly corresponds to the DIB at ${\lambda 6284}$ in the host galaxy rest frame. 
Upon closer inspection however, the feature does not align with the Doppler shift of the other detected absorption features, 
casting doubt on the identification.
Other known DIBs, such as $\lambda\lambda 5780$ and $5797$, which are believed to correlate 
well with reddening \citep{2013ApJ...779...38P,2015MNRAS.447..545B}, are not detected above the noise level.
Assuming full-width-half-maxima from SN~2001el for DIBs $\lambda\lambda 5780$ and $5797$ of $2.0$ and $0.7$~\AA,
the corresponding EWs must be $<150$ and $<50$~m\AA, respectively given the signal-to-noise ratio.

From the UV, optical and IR photometry of \abc we can estimate the reddening of the SN by
fitting the SED of SN~2011fe \citep[as in][]{2014ApJ...788L..21A,2015MNRAS.453.3300A}
to the data.
Aside from the unusually fast rise of \abc, 
the above SED fits the lightcurve and colours well.
For this reason, we avoid including the early 
photometry and the light-curves using the photometry between phases $p=-10$ -- $+40$ days.

The measured reddening is estimated to $E(B-V)=0.07\ (0.05)$~mag for $R_V=1.4\ (3.1)$, i.e.
the full range of $R_V=1.4$--$3.1$ values measured for \sneia \citep[e.g.][]{2015MNRAS.453.3300A}.  
The fit further determined the light-curve parameters $t_{B_{\rm max}}=57498.76$ and $s=1.074$, 
which define the B-band maximum brightness in MJD and the lightcurve stretch with respect to 
SN~20011fe \citep[see e.g.][for a discussion on lightcurve stretch]{2001ApJ...558..359G}, respectively.
All statistical uncertainties are subdominant, and the errors are dominated by the assumed value of $R_V$.
The fitted values of $E(B-V)$ correspond to an extinction of $A_V=0.1\ (0.2)$~mag.  

The extinction can also be estimated by taking advantage of the standard candle property of \sneia and 
comparing the measured peak brightness with the expected for the given redshift, $z=0.0234$, of NGC~5221.  
Using the \snia lightcurve fitter SALT2 \citep{2007AA...466...11G}, 
we find that $A_V=-0.03\pm0.04$ which is well within the 
expected intrinsic peak brightness dispersion of $0.1$~mag.

In the past, grey dust ejected from galaxies  
has been proposed \citep{1999ApJ...525..583A,1999ApJ...512L..19A,2002A&amp;A...384....1G} 
as an alternative explanation to the observations interpreted as
a cosmological constant driven acceleration of the universe \citep{1999ApJ...517..565P,1998AJ....116.1009R}.
Although quasar observations have shown that grey dust is not responsible for the dim appearance of cosmological
\snia samples \citep{2003JCAP...09..009M,2010MNRAS.406.1815M}, 
a small grey dust component could still lead to biased measurements of cosmological samples.
Matter streaming from a galaxy is a suggested location where grey dust could be forming.
In such a scenario the \los to \abc should probe an overdensity of grey dust, 
and the standard candle properties of this SN Ia can be used to constrain it. 
Since the peak brightness of \abc is within the intrinsic scatter of \sneia however,
the grey dust column along the \los must be negligible.

 \begin{table*}
  \centering
  \begin{tabular}{c r@{\,}l r@{\,}l c c r@{\,}l r@{\,}l c c}
    \hline\hline
    {\small v } & \multicolumn{2}{c}{\small \naid\/1} & \multicolumn{2}{c}{\small \naid\/2} & {\small ratio} &
    {\small N(\nai)} &
    \multicolumn{2}{c}{\small \caii H} & \multicolumn{2}{c}{\small \caii K} & {\small ratio} &
    {\small N(\caii)}\\
    {\small (km s$^{-1}$)} & \multicolumn{2}{c}{\small (m\AA)} & \multicolumn{2}{c}{\small (m\AA)} 
    & {\small (D2 / D1)}  & {\small ($\times10^{12}$ cm$^{-2}$)} &\multicolumn{2}{c}{\small (m\AA)} 
    & \multicolumn{2}{c}{\small (m\AA)}  & {\small (K / H)} & {\small ($\times10^{12}$ cm$^{-2}$)}\\ 
    \hline
	{\small -77} & {\small 241} & {\small $\pm$ 3} & {\small 279} & {\small $\pm$ 3} &
	{\small 1.16} & {\small $>2.4$} &
	{\small 42} & {\small $\pm$ 3} & {\small 73} & {\small $\pm$ 3} & {\small 1.74} & {\small 1.0 $\pm$ 0.1}\\
	{\small -51}  & {\small 278} & {\small $\pm$ 3} & {\small 312} & {\small $\pm$ 4} &
	{\small 1.12} & {\small $>2.8$}
	& {\small 110} & {\small $\pm$ 3} & {\small 166} & {\small $\pm$ 4} & {\small 1.51} & {\small 3.0 $\pm$ 0.3}\\
    \hline\hline
\multicolumn{10}{l}{}
  \end{tabular}
  \caption{Average \naid and \cahk equivalent widths of the features with centroids at 
  $-77$ and $-51$~km~s$^{-1}$ measured from the UVES spectra. 
  The equivalent width ratios are indicative of the optical depth of the absorbers.
  Column density lower limits are given for \nai, which were computed from \naid\/1 assuming
  the optically thin relation to equivalent width.
  The \caii column densities were determined from the 5 component Voigt profile fit. 
  \label{tab:cd}}
\end{table*}

\section{Absence of circumstellar absorption}
\label{sec:cs}

   \begin{figure}
   \centering
   	\resizebox{\hsize}{!}{\includegraphics{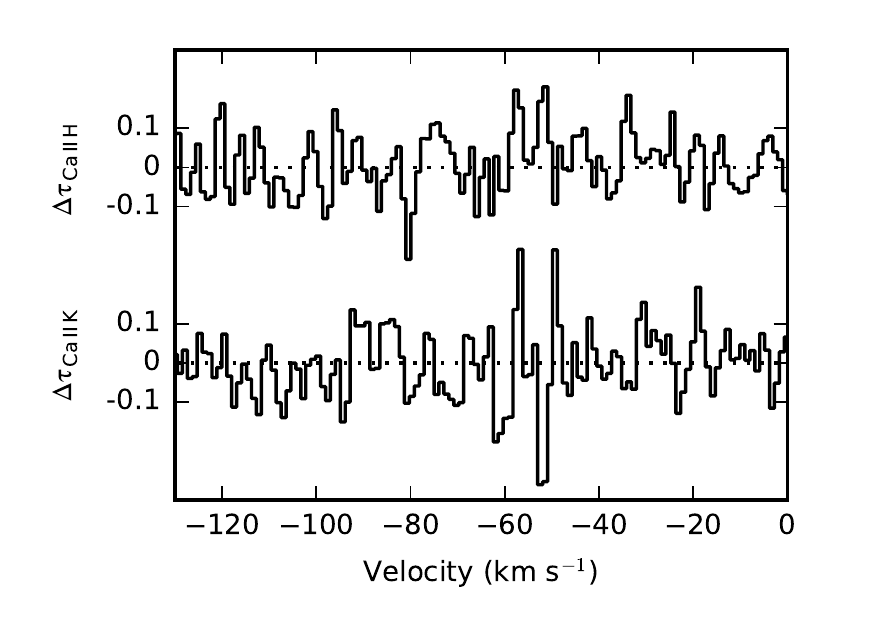}}
   \caption{Difference in apparent optical depth between the UVES spectra of \cahk. No pixels seem to be significant outliers.
 }
         \label{fig:tau}
    \end{figure}

Although the majority of the absorption lines detected in \snia spectra 
are believed to be due to gas in the ISM 
of their host galaxies, there exist observations that indicate the presence of CS gas.
For example, cases of slight variations in absorption line profiles 
\citep{2007Sci...317..924P,2009ApJ...702.1157S,2013A&amp;A...549A..62P,2015ApJ...801..136G,2016A&amp;A...592A..40F}
hint at photoionisation or recombination of absorbers
and predominantly blue-shifted \naid profiles \citep{2011Sci...333..856S,2013MNRAS.436..222M} 
suggest outflowing gas.

Nevertheless, the existence of CS gas surrounding \sneia remains uncertain. 
Most of the known cases of varying absorption lines all occurred in crowded fields,
where geometric effects \citep{2010A&A...514A..78P} might be expected 
and a larger sample of multi-epoch high-resolution spectra 
has not turned up any more examples \citep{2014MNRAS.443.1849S}.

\sneia peak in UV earlier than in optical. 
Because of this, most photoionisation should occur before the SNe reach maximum brightness
\citep{2009ApJ...699L..64B, 2016A&amp;A...592A..40F}.
Ideally, early spectra (around -14 days before maximum) need to be obtained to search for photoionisation
of CS gases.
The remote location and early discovery of \abc made it a good candidate \snia to search for CS matter 
by looking for changes in absorption line profiles.

Knowing the SED of \sneia as well as the ionisation energy and cross-section of an absorption species, 
the rate of photoionisation depends on the distance of the gas from the SN.
Thus measuring the ionisation rate, is a method to determine the distance of an absorber to the supernova.
Thereby the different ionisation energies and cross-sections of different absorption species implies 
that they are sensitive to photoionisation at different distances.

In the absence of any variations, one can determine the distances at which there was no detectable 
gas, which would have been been ionised.
The exclusion range is defined by two radii. The inner radius is determined by the distances at which all
gas is ionised before the first spectrum is taken. The outer radius is defined by the distance at which the amount 
of ionisation that occurs by the time the last spectrum is taken is negligible.  



In Table~\ref{tab:ew} we present the time series of the measured EWs of \naid and \cahk. 
It can be seen that there is neither significant time variability in the EW measurements, 
nor are there any visible changes in the line profiles.
We investigate the profiles of \cahk in the UVES spectrum in more detail than the saturated \naid lines.
A direct comparison of the Voigt profile parameters, reveals no difference between the two UVES epochs.
Furthermore, we plot the difference in the apparent optical depth of \cahk between the two epochs
in Figure~\ref{fig:tau}. We find no significant changes in the apparent optical depth of any part of the profiles.

\begin{figure}
   \centering
   	\resizebox{\hsize}{!}{\includegraphics{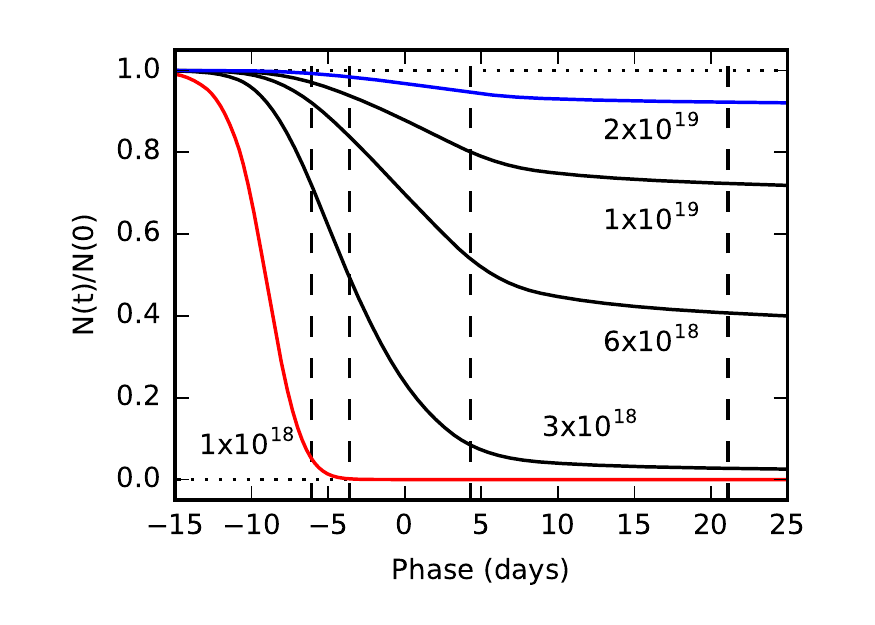}}
   \caption{Fractional photoionisation curves of \nai gas at annotated radii in cm. The vertical dashed lines indicate the 
   phases of the obtained spectra. The red curve at $1\times10^{18}$~cm, defines the inner exclusion radius at which 
   a gas cloud would have been ionised before the first spectrum was obtained. The blue curve defines the outer exclusion 
   radius at $2\times10^{19}$~cm, where photoionisation leads to negligible change.
 }
         \label{fig:ion}
    \end{figure}

Using the method of \citet{2016A&amp;A...592A..40F},
we can exclude the presence of \nai gas at a distance of $R^{\rm \,excl}_{\rm\nai}\approx1\times10^{18}$ -- $2\times10^{19}$~cm, 
and \caii gas at $R^{\rm \,excl}_{\rm\caii}\approx8\times10^{16}$ -- $3\times10^{17}$~cm 
from \abc at the $3\sigma$ confidence level.
To illustrate the exclusion range, Figure~\ref{fig:ion} shows \nai ionisation curves of gas clouds at the limiting  
and exclusion radii.
The exclusion limits do not include possible errors in the \sneia SED as described in \citet{2016A&amp;A...592A..40F}.
In particular, the exclusion limits obtained from \caii must be taken with caution, due to
the diversity of the \snia SEDs.

Thus all the detected absorption features must originate from gas farther from \abc than the outer exclusion limit
and are part of the ISM.
It is not possible to further constrain the relation of \abc with the gas clouds other than stating that 
the supernova must have occurred behind them.
Due to the broad nature of intrinsic supernova spectral features,
the systemic velocity of the supernova can't be determined with accuracy.
It is nevertheless plausible that the progenitor system of \abc was moving with the tidally stripped gas
behind which it exploded.

  \begin{table*}
  \centering
  \begin{tabular}{l c c c c r@{\,}l r@{\,}l r@{\,}l r@{\,}l r@{\,}l}
    \hline\hline
    {\small Instrument}  &  {\small Date} &  {\small Phase} & {\small $R$ $(\lambda/\delta \lambda)$} & {\small S/N} & 
    \multicolumn{2}{c}{\small \naid\/1} & \multicolumn{2}{c}{\small \naid\/2} & \multicolumn{2}{c}{\small \caii H}   &  \multicolumn{2}{c}{\small \caii K} \\ 
    & & {\small (days)} & & & \multicolumn{2}{c}{\small (m\AA)} & \multicolumn{2}{c}{\small (m\AA)} & \multicolumn{2}{c}{\small (m\AA)} & \multicolumn{2}{c}{\small (m\AA)} \\ 
    \hline
	{\small XShooter} & {\small Apr. 14} & {\small -6.1} & {\small 7,450$^{\dagger}$} & {\small 113} &
	{\small 512}& {\small $\pm$ 6} & {\small 606}& {\small $\pm$ 6} & {\small 141}&{\small $\pm$ 4} & {\small 238}& {\small $\pm$ 8} \\
	{\small UVES} & {\small Apr. 16} & {\small -3.6} & {\small 68,000} & {\small 20} &
	{\small 517}& {\small $\pm$ 7} & {\small 585}& {\small $\pm$ 7} & {\small 147}& {\small $\pm$ 7} & {\small 240}& {\small $\pm$ 6}\\
	{\small UVES} & {\small Apr. 24} & {\small 4.3} & {\small 87,000} & {\small 25} &
	{\small 511}& {\small $\pm$ 5} & {\small 597}& {\small $\pm$ 5} & {\small 155}& {\small $\pm$ 6} & {\small 238}& {\small $\pm$ 6} \\
	{\small XShooter} & {\small May 12} & {\small 21.2} & {\small 7,450$^{\dagger}$} & {\small 73} &
	{\small 497}& {\small $\pm$ 10} & {\small 592}& {\small $\pm$ 10} & {\small 138}& {\small $\pm$ 8} & {\small 227}& {\small $\pm$ 9} \\
	\hline
	& \multicolumn{4}{c}{\small Average} & {\small 511}&{\small $\pm$ 3} & {\small 597}&{\small $\pm$ 3} & {\small 145}&{\small $\pm$ 3} & {\small 237}&{\small $\pm$ 3} \\
    \hline\hline
    \multicolumn{5}{l}{\small {$^{\dagger}$nominal XShooter resolution}}\\
  \end{tabular}
  \caption{Equivalent widths of  Na~I~D, Ca~II~H\&K, of iPTF~16abc using XShooter and UVES. 
  	The resolution of the UVES spectra were estimated from the full width at half-maximum intensity (FWHM) 
	of several telluric features, and the S/N per pixel is measured around the wavelength of Na~I~D.
  \label{tab:ew}}
\end{table*}

 \section{Discussion} 
 \label{sec:host}

The position of \abc suggests that we are probing the ISM in the tidal arm of \host. 
\abc is $170"$ from the center of \host, which at a redshift of $z=0.0234$ \citep{2015MNRAS.447.1531C} 
 corresponds to at least $80$~kpc.

The absorption lines span a radial velocity of $-100$~--~$-30$~km~s$^{-1}$,
with respect to the rest-frame of the host galaxy core.
These values are smaller than the rotation velocity of \host.
\citet{2015MNRAS.447.1531C} quote a line width (twice the rotation velocity) 
of $510\pm20$~km~s$^{-1}$.
Furthermore, two Sloan Digital Sky Survey (SDSS) spectra of spiral arms on opposite 
sides of the galaxy show a velocity difference 
of $477$~km~s$^{-1}$. 
The coordinates at which the SDSS spectra have been taken prior to explosion, are marked in Figure~\ref{fig:field},
whereby the eastern arm (S1) is approaching and the western arm (S2) is 
receding with $-254$ and $223$~~km~s$^{-1}$, respectively.
Interestingly, this implies that the gas in the tidal tail is receding from the galaxy in the 
opposite direction to the spiral arm it is closest to.
Although the tidal tail projects in the opposite direction, 
this may indicate that the tidal stream connects to the eastern spiral arm.

\begin{figure}
   \centering
   	\resizebox{\hsize}{!}{\includegraphics{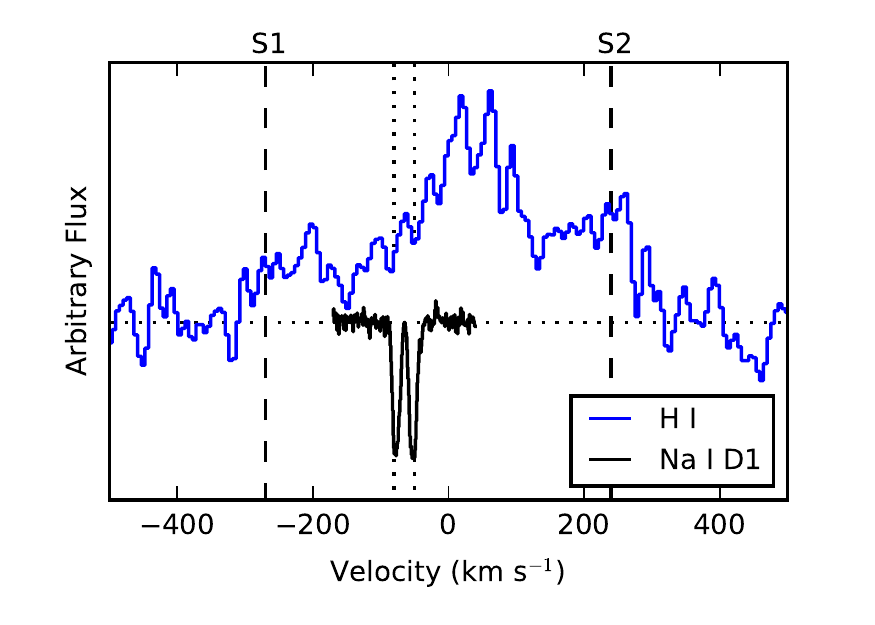}}
   \caption{Comparison of the radial velocity of the \nai gas along the \los of \abc and \hi in and around \host.
   The ALFA \hi spectrum encompasses the position of \abc and \host.
  The \naid\/1 profile is shown from the Apr. 24 UVES spectrum.
   Vertical dashed lines indicate the \los velocity of the two SDSS spectra with respect to the rest-frame of \host.
   The labels S1 and S2 corresponding to those in Figure~\ref{fig:field}.
 }
         \label{fig:hi}
    \end{figure}

To further compare the radial velocity of the gas seen in the \abc spectra with \host,
we consider an \hi profile by the Arecibo L-band Feed Array (ALFA) as part of the 
Arecibo Legacy Fast ALFA survey \citep[ALFALFA,][]{2005AJ....130.2598G,2011AJ....142..170H}.
The beam of the ALFA encompasses both the position of \abc and \host.
In Figure~\ref{fig:hi}, the \naid\/1 profile of the \abc\ \los is compared to the ALFA \hi profile of the entire galaxy.

The position of \abc 
allows us to estimate a time-frame within which the gas must have been stripped from \host. 
Assuming the gas is moving away from the galaxy with a velocity of $v\approx100$~km~s$^{-1}$, 
it would take at least $8\times10^8$~years to reach a distance of $80$~kpc, depending on the 
exact projection angle.
Models show that this time frame is consistent with close 
encounters and mergers of large galaxies \citep{1972ApJ...178..623T}.

 A number of galaxies, listed in Table~\ref{tab:gals}, are at a comparable redshift to \host 
 and within a few hundred kpc from it.
 The two most prominent galaxies in the list, NGC~5222 and NGC~5230, 
 are likely candidates to have interacted with \host, producing the observed tidal stream.
 \citet{2016AJ....152..225H} have studied NGC~5230 as an example of a very \hi rich galaxy.
 In fact, the ALFALFA data they present in their study show that \hi gas encompasses 
 the entire galaxy group, with column density maxima corresponding to \host, NGC~5222 and NGC~5230.
 Thus the gas along the \los of \abc is representative of gas
 that can end up in the CGM of the group via interactions between the galaxies. 
 
 Narrow absorption lines, such as those detected in the spectra of \abc, are not unusual features for 
 \snia spectra \citep[e.g. the sample of][]{2011Sci...333..856S}.
 However, these features typically appear in \loss of supernovae clearly situated within their host galaxies.
 The \nai and \caii gas columns along the \los of \abc appear to be  comparable to typical \loss within galaxies.
 To illustrate this, we consider the \naid absorption in the two SDSS spectra available of \host.
We find \naid EWs of $2.1\pm0.5$ and $2.9\pm0.9$~\AA\ for the eastern and western spectra respectively.
 \cahk lines are unfortunately buried in noise in the SDSS spectra and cannot be measured.
 
 Although the \nai columns in the tidal stream and the \host differ by close to an order of magnitude,
 it is important to note the differences between the SDSS and supernova \naid EWs.
In the case of the SDSS spectra, the absorption lines represent the average absorption along the \loss
to all stars covered by the spectral fibre.
The investigated column can thus be considered to be the average 
across an area of $\approx 6$~kpc$^2$ area of the galaxy. 
In comparison to this, the column measured in the supernova spectrum, 
spans the projected area of a photosphere which is of the order of $\approx10^{-6}$~pc$^2$ 
at maximum brightness\footnote{Approximated from $10^4$~km~s$^{-1}$ expansion after 21 days.}.
We are thus comparing a very small area in the tidal tail to the average of a large area within the galaxy, 
where many more gas clouds with different velocities can be situated along the \los.
 
 From ISM studies of the Milky Way, it is known that \naid and \cahk correlate with 
 reddening and also \hi column densities \citep{2012MNRAS.426.1465P,2015MNRAS.452..511M}.
 These empirical relations can be used to estimate $E(B-V)$ and $N({\rm \hi})$ for the \los of \abc.
 Using the saturated \naid lines, the \citet{2012MNRAS.426.1465P} relation suggests that 
 $E(B-V)_{\rm est}>0.32$.
 It is known that the empirical relations of \naid to reddening 
 work poorly for \sneia \citep{2011MNRAS.415L..81P,2013ApJ...779...38P}. 
 \abc is another example with deep \naid absorption lines, but comparatively little photometric reddening.
Recently, \citet{2017ApJ...836...13H} suggested that the unusually large \nai columns could 
 originate from gas released during dust grain collisions in clouds irradiated by the SNe.
 The ratio $N({\rm\nai})/N({\rm\caii})>1.3$ falls outside the range over which the \citet{2015MNRAS.452..511M} relation to 
 \hi has been determined.
 Extrapolating the relation would imply an
 \hi column $N(\rm \hi)_{\rm est.}>10^{21}$~cm$^{-2}$.
 
 The \hi column density estimates are more difficult to compare to.
 From the ALFA spectrum, we can compute that the entire beam, spanning $3.5'\times3.8'$
  contains M$_{\rm \hi}=1.0\times10^{10}$~M$_{\odot}$, corresponding to an average 
  column density of $N(\rm \hi)=3.7\times10^{19}$~cm$^{-2}$.
  Most of the \hi in the ALFA spectrum must be situated in \host, but could also have been stripped 
  to locations such as the \los of \abc.


\begin{table*}
       	\centering
  	\begin{tabular}{c c c c l}
    	\hline\hline
	{\small Galaxy} & {\small Redshift} & {\small perp. Dist.$^{\dagger}$} & {\small Direction} & {\small Morphology / Notes} \\
	& & {\small (kpc)} & & \\
    	\hline
	{\small 2MASX J13341283+1349334} & {\small 0.0235}& {\small 296} & {\small W} & {\small dwarf}\\
	{\small 2MASX J13343006+1348154} & {\small 0.0242}& {\small 184} & {\small WSW} & {\small ?}\\
	{\small SDSS J133443.53+134507.3} & {\small 0.0253}& {\small 161} & {\small SW} & {\small ? elongated dwarf}\\
	{\small 2MASX J13344815+1344391} & {\small 0.0222}& {\small 159} & {\small SSW} & {\small dwarf}\\
	{\small NGC~5222} & {\small 0.0231}& {\small 154} & {\small S} & {\small large elliptical}\\
	{\small -- } & {\small 0.0228} & {\small 150} & {\small S} &{\small irregular overlapping merging with NGC~5222 } \\
	{\small NGC~5230} & {\small 0.0229}& {\small 363} & {\small SE} & {\small large face-on spiral}\\
	{\small SDSS J133509.69+135324.0} & {\small 0.0234}& {\small 136} & {\small NE} & {\small dwarf}\\
	{\small NGC~5226} & {\small 0.0244}& {\small 161} & {\small NNE} & {\small spiral?}\\
	\hline\hline
	\multicolumn{3}{l}{\small $^{\dagger}$At z$=0.0234$.}\\
  	\end{tabular}
  	\caption{Neighbouring galaxies of \host at comparable redshift.
	\label{tab:gals}}
\end{table*}

\section{Summary and conclusions}
\label{sec:con}
We have presented the interstellar absorption lines observed 
in the spectra of Type Ia supernova \abc.
The gas corresponds to the tidally stripped ISM at least $80$~kpc from the centre of the host galaxy \host.
Compared with a \los in a typical galaxy, the ISM in the tidal tail  
appears to have a typical gas content, but a surprisingly small column of dust.

We detected \naid and \cahk absorption features in two distinct clusters 
at velocities $-77$ and $-51$~km~s$^{-1}$  from the systemic velocity of \host.

From the lack of variations in the  \naid and \cahk profiles, we determined that the 
observed gas cannot be part of the CS environment of the supernova, since 
photoionisation would have resulted in a significant change in the column density.
Thus the gas seen along the \los of \abc is the tidally stripped ISM of \host.

At the same time the standard candle nature of \abc has allowed us to determine
that the supernova is barely reddened compared with the normal \sneia colours,
implying that there is a negligible amount of dust along the \los.
This further excludes the presence of grey dust,
which could be along the \los of \abc,
if grey dust existed in the inter-galactic medium.

\host likely had a close encounter with one of its neighbouring galaxies 
in the past $\approx10^9$ years, when large portions of gas and the progenitor system 
of \abc were tidally stripped from it. 
While the velocity of the gas in the tidal arm appears comparable to the velocity of the eastern spiral arm, 
the projection of the tail points in the opposite direction.
This suggests that the tidal tail might be connected to the eastern spiral arm.

A map of \hi content of the group of galaxies including \host, presented by \citet{2016AJ....152..225H},
suggests that the group shares a common gas envelope.
As the galaxies within the group are likely to continue to interact and merge in the future,
the gas seen along the \los of \abc is representative of the gas that is transplanted into the 
CGM surrounding the galaxy group.
It has been shown that dense cold gas clouds can exist in the CGM of galaxy groups 
for $>5\times10^{8}$~years \citep{2015ApJ...812...78B}.
In future interactions with neighbouring galaxies, these clouds can be the sites of star formation.
In the long run however, the gas clouds could dissipate, enriching the CGM with the detected metals.

The presented gas profiles along the \los of \abc should be useful to future studies of \host, its tidal tail 
and the galaxy group it is located in.
In particular, the profiles could complement higher resolution observations of the \hi gas in \host and the tidal arm, 
resolving the dynamics of the galaxy group.

\begin{acknowledgements}

We thank Gregory Hallenbeck for his help with extracting ALFALFA data, David Martinez-Delgado, Darach Watson 
and Brice Menard for their helpful insights and Jesper Sollerman and Avishay Gal-Yam for their comments.
R.A. and A.G. acknowledge support from the Swedish Research Council and the Swedish Space Board. 
The Oskar Klein Centre is funded by the Swedish Research Council.
This work is based on observations made with the Nordic Optical Telescope, operated by the Nordic Optical Telescope Scientific Association at 
the Observatorio del Roque de los Muchachos, La Palma, Spain, of the Instituto de Astrofisica de Canarias.
This work makes use of observations from the LCOGT network.
The Pan-STARRS1 Surveys (PS1) and the PS1 public science archive have been made possible through contributions by the Institute for Astronomy, the University of Hawaii, the Pan-STARRS Project Office, the Max-Planck Society and its participating institutes, the Max Planck Institute for Astronomy, Heidelberg and the Max Planck Institute for Extraterrestrial Physics, Garching, The Johns Hopkins University, Durham University, the University of Edinburgh, the Queen's University Belfast, the Harvard-Smithsonian Center for Astrophysics, the Las Cumbres Observatory Global Telescope Network Incorporated, the National Central University of Taiwan, the Space Telescope Science Institute, the National Aeronautics and Space Administration under Grant No. NNX08AR22G issued through the Planetary Science Division of the NASA Science Mission Directorate, the National Science Foundation Grant No. AST-1238877, the University of Maryland, Eotvos Lorand University (ELTE), the Los Alamos National Laboratory, and the Gordon and Betty Moore Foundation.

\end{acknowledgements}

%
%

\bibliographystyle{aa} 
\bibliography{abc} 


\Online

\end{document}